\documentclass[prb,twocolumn,byrevtex]{revtex4}%
\usepackage{amsfonts}
\usepackage{amsmath}
\usepackage{amssymb}
\usepackage{graphicx}%
\begin{document}
\title{Spin precession due to spin-orbit coupling in a two-dimensional electron gas
with spin injection via ideal quantum point contact}
\author{Ming-Hao Liu}
\affiliation{Department of Physics, National Taiwan University, Taipei, Taiwan}
\author{Ching-Ray Chang}
\affiliation{Department of Physics, National Taiwan University, Taipei, Taiwan}
\author{Son-Hsien Chen}
\affiliation{Department of Physics, National Taiwan University, Taipei, Taiwan}

\begin{abstract}
We present the analytical result of the expectation value of spin resulting
from an injected spin polarized electron into a semi-infinitely extended 2DEG
plane with [001] growth geometry via ideal quantum point contact. Both the
Rashba and Dresselhaus spin-orbit couplings are taken into account. A
pictorial interpretation of the spin precession along certain transport
directions is given. The spin precession due to the Rashba term is found to be
especially interesting since it behaves simply like a windshield wiper which
is very different from the ordinary precession while that due to the
Dresselhaus term is shown to be crystallographic-direction-dependent. Some
crystallographic directions with interesting and handleable behavior of spin
precession are found and may imply certain applicability in spintronic devices.

\end{abstract}
\maketitle

Recent research on spin-polarized electron transport in semiconductors has
attracted a great attention in the emerging field of spintronics.\cite{SM
spintronics} Of particular interests is the manipulation of spin via
spin-orbit (SO) coupling in semiconductor nanostructures. In a two-dimensional
electron gas (2DEG) confined in a heterostructure quantum well (QW), two basic
mechanisms of the SO coupling are often taken into account: (i) structure
inversion asymmetry (SIA) mechanism described by the Rashba term,\cite{Rashba
term}%
\begin{equation}
H_{R}=\frac{\alpha}{\hbar}\left(  p_{x}\sigma_{y}-p_{y}\sigma_{x}\right)  ,
\label{HR}%
\end{equation}
whose coupling strength $\alpha$ is
gate-voltage-dependent,\cite{DasB1989,LuoJ1990,NittaJ1997,EngelsG1997,HeidaJP1998,HuCM1999,GrundlerD2000}
and (ii) bulk inversion asymmetry (BIA) mechanism described by the Dresselhaus
term.\cite{Dresselhaus term,LommerG1985} When restricted to a two-dimensional
semiconductor nanostructure with [001] growth geometry, this term is of the
form\cite{DyakonovMI1986,BastardG1992}%
\begin{equation}
H_{D}=\frac{\beta}{\hbar}\left(  p_{x}\sigma_{x}-p_{y}\sigma_{y}\right)
\label{HD}%
\end{equation}
where the coupling parameter $\beta$ is material specific. The interface
inversion asymmetry (IIA)\cite{VervoortL1997,RosslerU2002} also provides
certain contribution to the Dresselhaus term in the SO coupling but is
phenomenologically inseparable from BIA.

Whereas the competition between Rashba and Dresselhaus terms was concluded
that the former dominates in narrow-gap systems\cite{DasB1989,LuoJ1990} while
the latter dominates in wide-gap materials,\cite{LommerG1988} Datta and Das
proposed an theoretical idea constructing an electronic analog of the optic
modulator using ferromagnetic contacts as spin injector and detector with a
2DEG channel confined in a narrow-gap semiconductor with only the Rashba SO
coupling taken into account.\cite{Datta-Das} In their proposal, the spin
precession is envisioned due to the interference between the two
eigenfunctions superposing the wave function of the injected spin with a
gate-voltage-tunable phase difference $\Delta\theta=2m^{\ast}\alpha
L/\hbar^{2}$ with $L$ being the channel length. Therefore, the spin
orientation angle for electrons arriving at the end of the 2DEG channel, and
hence the resulting current, is theoretically tunable via the applied
gate-voltage. Hence the proposed device is expected to serve as a field-effect
transistor (FET) based on the electron spin and has been commonly referred to
as the Datta-Das spin-FET.

Recently, Winkler has further demonstrated this well-known spin precession
described above and also the spin orientation in a quasi-two-dimensional
(quasi-2D) electron system by using an $8\times8$ Kane model which takes into
account both SIA and BIA mechanisms.\cite{Winkler} The spin orientation is
shown to be sensitively dependent on the crystallographic direction for which
the quasi-2D system is grown. This is also consistent with the previous
results obtained by \L usakowski \emph{et al.} showing that the conductance of
the Datta-Das spin-FET depends significantly on the crystallographic direction
of the channel when the Dresselhaus term is also at
present.\cite{LusakowskiA2003} Indeed, the contribution to the SO coupling of
Rashba and Dresselhaus terms may be of the same order in some QWs (such as
GaAs QWs\cite{JusserandB1995}) and their ratio is even shown to be
experimentally determinable very recently.\cite{GanichevSD2004} Therefore,
possible effects caused by the Dresselhaus term on spin-related devices has
been an imperative issue in semiconductor
spintronics.\cite{LusakowskiA2003,TingDZY}

In this report, we extend Winkler's work,\cite{Winkler} who calculated the
expectation value of the spin operator $\left\langle \mathbf{S}\right\rangle $
(referred to as the spin vector in his paper) with respect to the injected
spin-polarized electron state superposed by the two eigenstates of the 2DEG in
the presence of\ the SO coupling. Whereas the spin precession is shown by
calculating the overlaps between the spin vector and the polarization of the
ferromagnetic drain contact numerically, we present the analytical result of
the spin vector as a function of the coupling strengths $\alpha$ and $\beta,$
the orientation angle of the injected spin, and the position of determination.
A pictorial interpretation of the spin precession along certain transport
directions is given. By analyzing the two extreme cases, pure Rashba and pure
Dresselhaus, the spin precession due to SO couplings in inversion-asymmetric
2DEGs can be understood more concretely. Some crystallographic directions with
interesting and handleable spin precession behavior are found and may imply
certain applicability in spintronics.

Consider an electron with definite spin perfectly injected from a
spin-polarized needle into an inversion-asymmetric 2DEG where both the Rashba
and Dresselhaus SO couplings are present. The spin-injector and the 2DEG is
connected with an ideal quantum point contact and the 2DEG is assumed to be
semi-infinitely extended so that the boundary effect is out of consideration.
Let the electron injected at an angle $\phi$ with spin $\mathbf{S}_{0}$
orienting toward $\phi_{s}$ with respect to $x$ axis. Setting the growth
direction of the 2DEG layer to be [001], and the $x$ and $y$ axes to be [100]
and [010], respectively, the single electron Hamiltonian under the effective
mass approximation can be written as%
\begin{equation}
H=\frac{p^{2}}{2m^{\ast}}\sigma_{0}+H_{R}+H_{D}%
\end{equation}
where $m^{\ast}$ is the electron effective mass in the 2DEG. Defining%
\begin{equation}
\gamma\left(  \phi\right)  \equiv\sqrt{\alpha^{2}+\beta^{2}+2\alpha\beta
\sin2\phi} \label{gamma}%
\end{equation}
and
\begin{equation}
e^{-i\varphi}\equiv\frac{\alpha e^{-i\phi}-i\beta e^{i\phi}}{\gamma\left(
\phi\right)  }\text{,} \label{varphi}%
\end{equation}
the corresponding eigenenergies and eigenfunctions can be easily obtained as%
\begin{equation}
E_{\pm}=\frac{\left(  \hbar k_{\parallel}^{\pm}\right)  ^{2}}{2m^{\ast}}%
\pm\gamma\left(  \phi\right)  k_{\parallel}^{\pm} \label{eigenvalues}%
\end{equation}
and%
\begin{equation}
\left\langle \mathbf{r}|\mathbf{k}_{\parallel}^{\pm},\pm\right\rangle
\equiv\frac{1}{\sqrt{2}}e^{i\mathbf{k}_{\parallel}^{\pm}\cdot\mathbf{r}%
}\left(
\begin{array}
[c]{c}%
ie^{-i\varphi}\\
\pm1
\end{array}
\right)  \text{,} \label{eigenfunction (basis)}%
\end{equation}
where the inplane wave vector $\mathbf{k}_{\parallel}$ and the position vector
$\mathbf{r}$ represent two-dimensional vectors $\left(  k_{x},k_{y}\right)  $
and $\left(  x,y\right)  $, respectively. Separating the spin part from the
state kets $\left\vert \mathbf{k}_{\parallel}^{\pm},\pm\right\rangle $, we
denote the eigenspinors as $\left\vert \varphi-\pi/2,\pm\right\rangle $ with
the usual definition%
\begin{equation}
\left\vert \widetilde{\alpha},\pm\right\rangle \equiv\left\vert \widetilde
{\beta}=0,\widetilde{\alpha},\pm\right\rangle =\frac{1}{\sqrt{2}}\left(
\begin{array}
[c]{c}%
e^{-i\widetilde{\alpha}}\\
\pm1
\end{array}
\right)
\end{equation}
where $\widetilde{\beta}$ and $\widetilde{\alpha}$ are the polar and azimuthal
angles, respectively.\cite{Sakurai} Taking Eq. (\ref{eigenfunction (basis)})
as the basis, the injected spin $\left\vert \phi_{s},+\right\rangle $ can be
expanded as%
\begin{equation}
\left\vert \phi_{s},+\right\rangle =c_{+}\left\vert \varphi-\pi
/2,+\right\rangle +c_{-}\left\vert \varphi-\pi/2,-\right\rangle
\end{equation}
where%
\begin{align}
c_{\pm}  &  =\left\langle \varphi-\pi/2,\pm|\phi_{s},+\right\rangle
\nonumber\\
&  =\frac{1}{2}\left(  \allowbreak-ie^{i\left(  \varphi-\phi_{s}\right)  }%
\pm1\right)  \text{.}%
\end{align}
Since there is a phase difference $\Delta\theta\left(  \mathbf{r}\right)
=2m^{\ast}\gamma\left(  \phi\right)  r/\hbar^{2}$ between $\left\vert
\varphi-\pi/2,+\right\rangle $ and $\left\vert \varphi-\pi/2,-\right\rangle $,
the spin state ket at position $\mathbf{r}$ can be equivalently written as%
\begin{equation}
\left\vert \phi_{s},+\right\rangle _{\mathbf{r}}^{RD}=c_{+}e^{-i\frac
{\Delta\theta\left(  \mathbf{r}\right)  }{2}}\left\vert \varphi-\pi
/2,+\right\rangle +c_{-}e^{i\frac{\Delta\theta\left(  \mathbf{r}\right)  }{2}%
}\left\vert \varphi-\pi/2,-\right\rangle
\end{equation}
where the superscript "RD" denotes that both the Rashba and Dresselhaus terms
are nonvanishing. By computing the expectation values of $\mathbf{S}$ with
respect to the state ket $\left\vert \phi_{s},+\right\rangle _{\mathbf{r}%
}^{RD}$, regardless of the factor $\hbar/2,$ we obtain%
\begin{align}
\left\langle \mathbf{S}\right\rangle _{\mathbf{r}}^{RD}  &  \equiv\left(
\begin{array}
[c]{c}%
\left\langle S_{x}\right\rangle _{\mathbf{r}}^{RD}\\
\left\langle S_{y}\right\rangle _{\mathbf{r}}^{RD}%
\end{array}
\right) \nonumber\\
&  =\left(
\begin{array}
[c]{c}%
\cos\phi_{s}\cos^{2}\frac{\Delta\theta}{2}-\cos\left(  2\varphi-\phi
_{s}\right)  \sin^{2}\frac{\Delta\theta}{2}\\
\sin\phi_{s}\cos^{2}\frac{\Delta\theta}{2}-\sin\left(  2\varphi-\phi
_{s}\right)  \sin^{2}\frac{\Delta\theta}{2}%
\end{array}
\right)  . \label{<S>}%
\end{align}
Note that the phase difference $\Delta\theta$ is generally a function of
$\mathbf{r}$ in the presence of both the Rashba and Dresselhaus terms, and
returns to be a constant only when either $\alpha$ or $\beta$ vanishes as will
be seen in the later discussion. In the following we discuss the behavior of
the spin precession under two extreme cases: $\left(  \alpha\neq
0,\beta=0\right)  $ and $\left(  \alpha=0,\beta\neq0\right)  $.

In the absence of the Dresselhaus term, we return to the familiar results of
the Rashba case: $\gamma\left(  \beta=0\right)  =\alpha$. Here the reason we
set Eq. (\ref{varphi}) can also be seen since the mathematical forms of the
eigenvalues, eigenfunctions, and the expectation values of $\mathbf{S}$
expressed in Eqs. (\ref{eigenvalues}), (\ref{eigenfunction (basis)}), and
(\ref{<S>}) are maintained. The results are obtained simply by replacing
$\varphi$ with $\phi$, e.g., the expectation value of $\mathbf{S}$ is%
\begin{equation}
\left\langle \mathbf{S}\right\rangle _{\mathbf{r}}^{R}=\left\langle
\mathbf{S}\right\rangle _{\mathbf{r}}^{RD}|_{\varphi=\phi}%
\end{equation}
where the superscript "R" denotes, similar to the previous ones, that only the
Rashba term is nonvanishing.%
\begin{figure}
[ptb]
\begin{center}
\includegraphics[
trim=1.929487in 2.364838in 1.697098in 3.179787in,
height=3.614in,
width=2.7103in
]%
{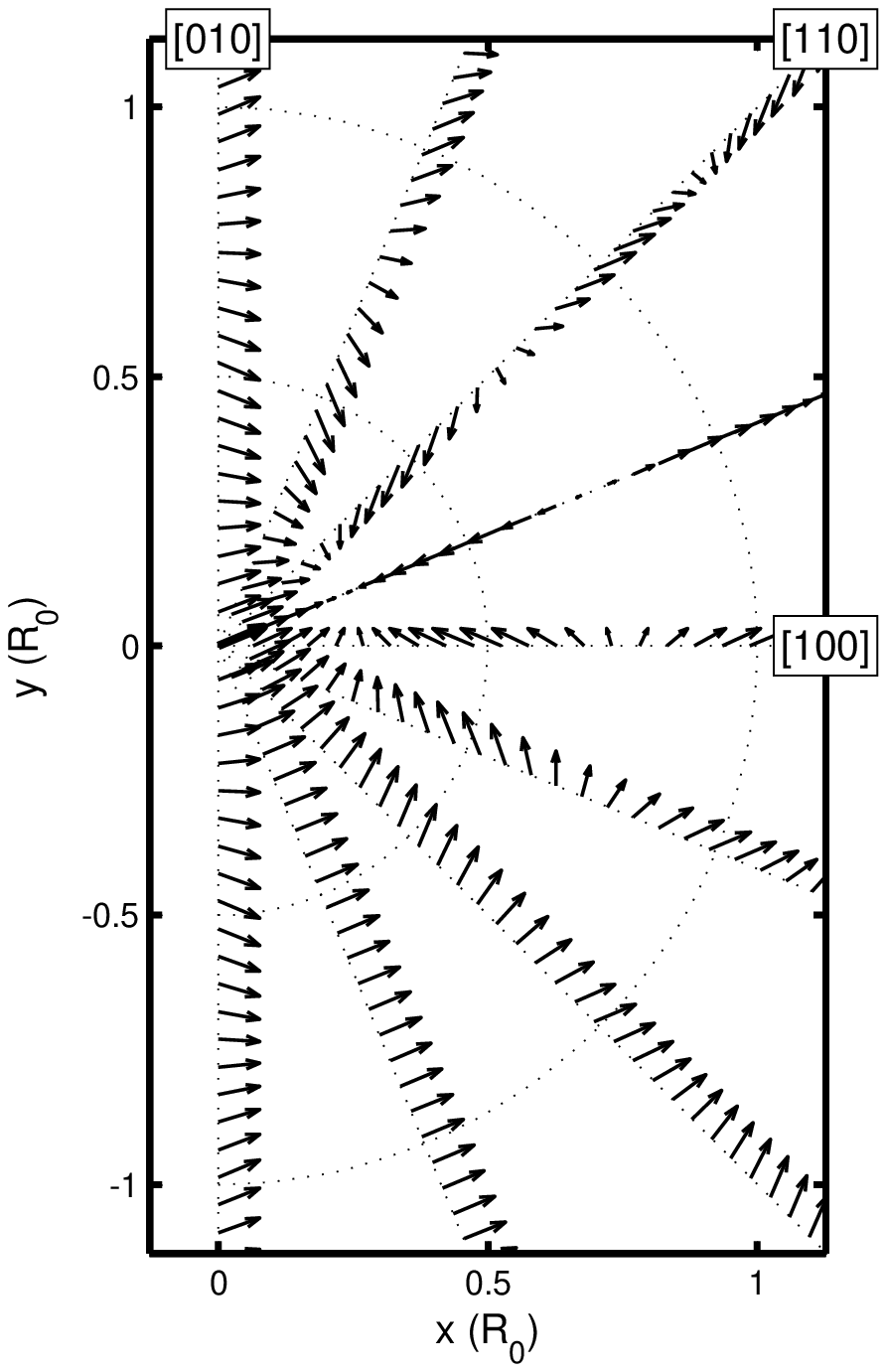}%
\caption{Spin precession due to the Rashba spin-orbit coupling in a 2DEG. Each
arrow indicates $\left\langle \mathbf{S}\right\rangle _{\mathbf{r}}^{R}$ on
the corresponding space point. The injected spin, shown by the bold arrow on
$\left(  0,0\right)  $, is set to orient $\pi/8$ with respect to [100]. The
compact unit $R_{0}$ is defined in Eq. (\ref{R0}) and the dotted lines are for
the guide of eyes.}%
\label{RSPfig}%
\end{center}
\end{figure}

As an example illustrating the Rashba spin precession, let $\phi_{s}\ $be
$\pi/8$. Spin orientations along 9 straight paths are shown in Fig.
\ref{RSPfig} where a compact unit defined as
\begin{equation}
R_{0}\equiv2\pi\hbar^{2}/(m^{\ast}\sqrt{\alpha^{2}+\beta^{2}}) \label{R0}%
\end{equation}
is used. Note that $R_{0}$ is essentially the length, within which the spin
completes one period of precession on $x$ or $y$ axes, and is typically of the
order of or less than $1$ $\mu$m for the Rashba case.\cite{Datta-Das} Each
pair of adjacent paths includes an angle of $\pi/8,$ which divides a half
circumferential angle into 8 equal parts. Spin precessions are clearly
observed except for the path which is perpendicular to the injected spin (see
the $-3\pi/8$ path in Fig. \ref{RSPfig}). This is reasonably expected since
the projection of the injected spin on one of the two eigen spin states, which
are always perpendicular to the electron wave vector, vanishes. Thus the fact
that only one component of the basis is occupied leads to zero spin precession.

Interestingly, the Rashba spin precession (RSP) behaves simply like a
windshield wiper swingling about the direction perpendicular to the electron
wave vector (or the propagation path). This is very different from the
ordinary full-circle precession. Another feature of the RSP is that the
projection of $\left\langle \mathbf{S}\right\rangle _{\mathbf{r}}^{R}$ on the
direction perpendicular to the path is universally conserved. Mathematically,
this can be further demonstrated by calculating $\left\langle \mathbf{S}%
\right\rangle _{\mathbf{r}}^{R}\cdot\hat{r}_{\parallel}$ and $\left\langle
\mathbf{S}\right\rangle _{\mathbf{r}}^{R}\cdot\hat{r}_{\perp}$ where $\hat
{r}_{\parallel}$ and $\hat{r}_{\perp}$ are the unit vectors in the directions
parallel and perpendicular to the path, respectively. Let us define these two
projection quantities to be $\left\langle \mathbf{S}\right\rangle
_{\mathbf{r,}\parallel}^{R}$ and $\left\langle \mathbf{S}\right\rangle
_{\mathbf{r,}\perp}^{R}.$ After some straightforward mathematical
manipulation, we obtain
\begin{subequations}
\label{<S>R}%
\begin{align}
\left\langle \mathbf{S}\right\rangle _{\mathbf{r},\parallel}^{R}{}  &
=S_{0,\parallel}\cos\Delta\theta\label{<S>R para}\\
\left\langle \mathbf{S}\right\rangle _{\mathbf{r},\perp}^{R}{}  &
=S_{0,\perp} \label{<S>R perp}%
\end{align}
where $S_{0,\parallel}=\cos\phi_{s}$ and $S_{0,\perp}=S_{0}\sin\phi_{s}$ are
the projections of the injected spin with normalized magnitude $S_{0}=1$ on
$\hat{r}_{\parallel}$ and $\hat{r}_{\perp},$ respectively.

Also, the transport directions along which no spin precession occurs can be
mathematically testified by calculating the scalar product of $\left\langle
\mathbf{S}\right\rangle _{\mathbf{r}}^{R}$ and $\mathbf{S}_{0}.$ The result is%
\end{subequations}
\begin{equation}
\left\langle \mathbf{S}\right\rangle _{\mathbf{r}}^{R}\cdot\mathbf{S}_{0}%
=\cos^{2}\frac{\Delta\theta}{2}-\cos\left[  2\left(  \phi-\phi_{0}\right)
\right]  \sin^{2}\frac{\Delta\theta}{2}%
\end{equation}
having the maximum value when $\phi=\phi_{0}+\left(  n+1/2\right)  \pi$ with
$n$ being an integer. That is, on the directions perpendicular to the injected
spin, we always have $\left\langle \mathbf{S}\right\rangle _{\mathbf{r}}%
^{R}=\mathbf{S}_{0}$ (see the $-3\pi/8$ path in Fig. \ref{RSPfig}).

\begin{figure}
[ptb]
\begin{center}
\includegraphics[
trim=1.923678in 2.404292in 1.704458in 3.140334in,
height=3.614in,
width=2.7103in
]%
{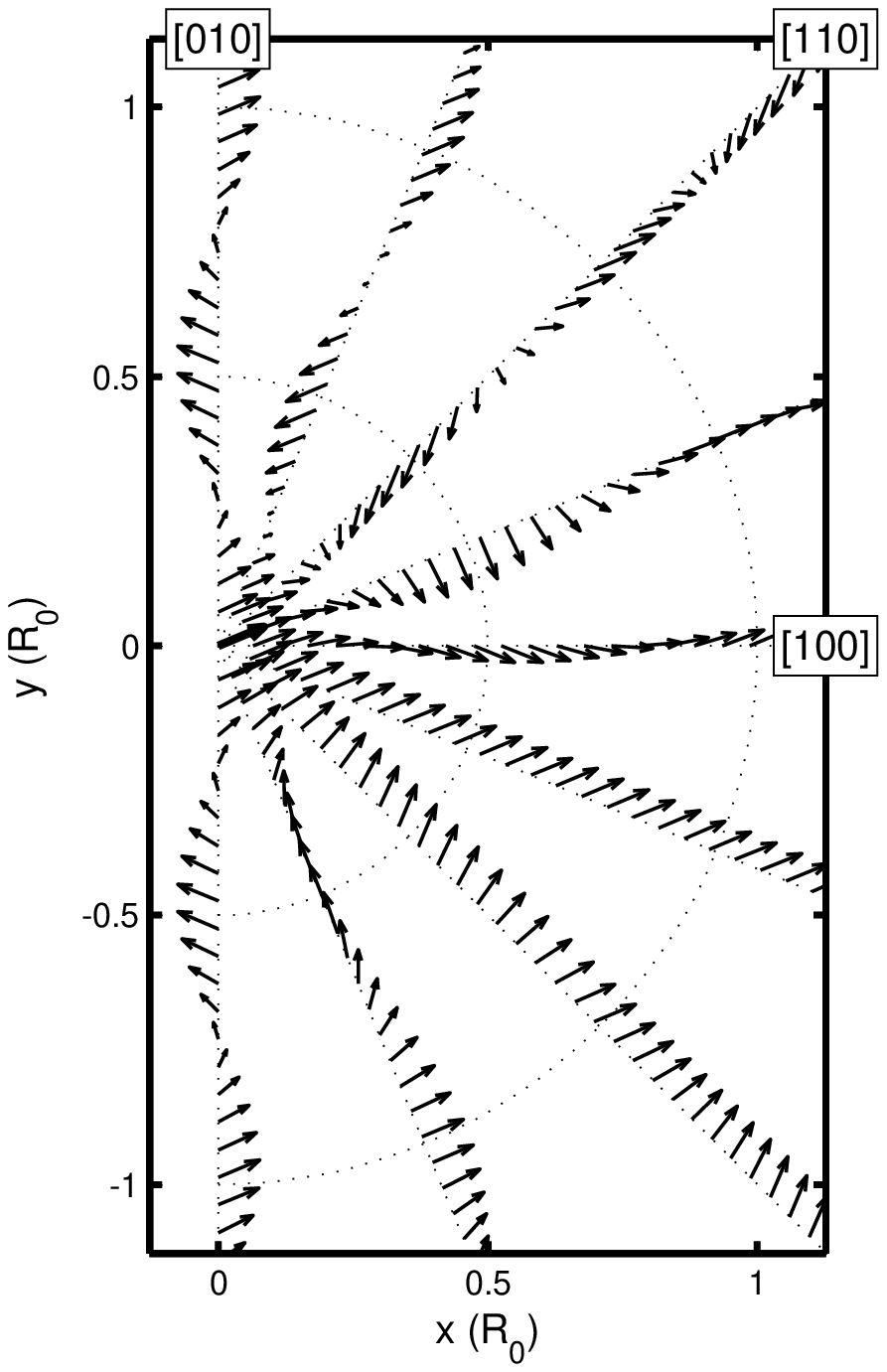}%
\caption{Spin precession due to the Dresselhaus spin-orbit coupling in a 2DEG.
Each arrow indicates $\left\langle \mathbf{S}\right\rangle _{\mathbf{r}}^{D}$
on the corresponding space point. The injected spin, shown by the bold arrow
on $\left(  0,0\right)  $, is set to orient $\pi/8$ with respect to [100]. The
compact unit $R_{0}$ is defined in Eq. (\ref{R0}) and the dotted lines are for
the guide of eyes.}%
\label{DSPfig}%
\end{center}
\end{figure}

In the absence of the Rashba term, we have $\gamma\left(  \alpha=0\right)
=\beta$ and $\varphi=\pi/2-\phi$ from Eqs. (\ref{gamma}) and (\ref{varphi}).
Thus the expectation value obtained in Eq. (\ref{<S>}) is modified as%
\begin{equation}
\left\langle \mathbf{S}\right\rangle _{\mathbf{r}}^{D}=\left(
\begin{array}
[c]{c}%
\cos\phi_{s}\cos^{2}\frac{\Delta\theta}{2}+\cos\left(  2\phi+\phi_{s}\right)
\sin^{2}\frac{\Delta\theta}{2}\\
\sin\phi_{s}\cos^{2}\frac{\Delta\theta}{2}-\sin\left(  2\phi+\phi_{s}\right)
\sin^{2}\frac{\Delta\theta}{2}%
\end{array}
\right)  \label{<S>D}%
\end{equation}
where the superscript "D" is, again, for reminding that only the Dresselhaus
term is at present. Again, the spin orientations on 9 straight paths are
plotted in Fig. \ref{DSPfig}.
The Dresselhaus spin precession (DSP), though appears to be more complicated
than the Rashba case, is still analyzable mathematically. We first turn to the
projections of $\left\langle \mathbf{S}\right\rangle _{\mathbf{r}}^{D}$ on
$\hat{r}_{\parallel}$ and $\hat{r}_{\perp}$. Using Eq. (\ref{<S>D}), we obtain%
\begin{align}
\left\langle \mathbf{S}\right\rangle _{\mathbf{r},\parallel}^{D}{} &
=S_{0,\parallel}\left(  \cos^{2}\frac{\Delta\theta}{2}+\cos4\phi\sin^{2}%
\frac{\Delta\theta}{2}\right)  \nonumber\\
&  -S_{0,\perp}\sin4\phi\sin^{2}\frac{\Delta\theta}{2}\label{<S>D para}%
\end{align}
and%
\begin{align}
\left\langle \mathbf{S}\right\rangle _{\mathbf{r},\perp}^{D}{} &  =S_{0,\perp
}\left(  \cos^{2}\frac{\Delta\theta}{2}-\cos4\phi\sin^{2}\frac{\Delta\theta
}{2}\right)  \nonumber\\
&  -S_{0,\parallel}\sin4\phi\sin^{2}\frac{\Delta\theta}{2}\text{.}%
\label{<S>D perp}%
\end{align}
The projections shown above in general do not exhibit conserved quantities as
in the Rashba case except two sets of paths: (i) For $\phi=n/4$ with $n$ being
an odd integer, we have
\begin{subequations}
\begin{align}
\left\langle \mathbf{S}\right\rangle _{\mathbf{r},\parallel}^{D} &
=S_{0,\parallel}\cos\Delta\theta\\
\left\langle \mathbf{S}\right\rangle _{\mathbf{r},\perp}^{D} &  =S_{0,\perp}%
\end{align}
which is exactly the same as Eq. (\ref{<S>R}). Thus on these directions,
namely, [110], [\={1}10], [\={1}\={1}0], and [1\={1}0] (for which only [110]
and [1\={1}0] are shown in our case) the spin precession behaves like the RSP
(see Fig. \ref{DSPfig}). (ii) For $\phi=n\pi/4$ with $n$ being an odd integer,
we have
\end{subequations}
\begin{subequations}
\begin{align}
\left\langle \mathbf{S}\right\rangle _{\mathbf{r},\parallel}^{D} &
=S_{0,\parallel}\\
\left\langle \mathbf{S}\right\rangle _{\mathbf{r},\perp}^{D} &  =S_{0,\perp
}\cos\Delta\theta
\end{align}
which is symmetric to the Rashba case such that the projection of
$\left\langle \mathbf{S}\right\rangle _{\mathbf{r}}^{D}$ on $\hat
{r}_{\parallel}$ being conserved while that on $\hat{r}_{\perp}$ being
oscillatory leads to another type of "swinging" spin precession on those
directions, namely, [100], [010], [\={1}00] (which is not shown in our case),
and [0\={1}0] (see Fig. \ref{DSPfig}). Unlike the Rashba case where the eigen
spin states are symmetric under rotation about [001] and the spin precession
depends only on the injected spin, these eight directions found above show
that there is intrinsic dependence of the DSP on crystallographic directions.

Similar to the Rashba case, there also exists one straight path on which no
spin precession occurs. This can be found, again, by calculating the scalar
product of $\left\langle \mathbf{S}\right\rangle _{\mathbf{r}}^{D}$ and
$\mathbf{S}_{0}$. Using Eq. (\ref{<S>D}) we obtain%
\end{subequations}
\begin{equation}
\left\langle \mathbf{S}\right\rangle _{\mathbf{r}}^{D}\cdot\mathbf{S}_{0}%
=\cos^{2}\frac{\Delta\theta}{2}+\cos\left[  2\left(  \phi+\phi_{0}\right)
\right]  \sin^{2}\frac{\Delta\theta}{2}%
\end{equation}
showing that for $\phi=-\phi_{0}+n\pi$ the spin precession vanishes (see the
$-\pi/8$ path in Fig. \ref{DSPfig}). Contrary to the Rashba case, this zero
spin precession path seems more like a mathematical, rather than geometrical,
result and thus reveals quite interesting physics.

In conclusion, we have presented the analytical results of the space-dependent
expectation values of the spin operator $\left\langle \mathbf{S}\right\rangle
_{\mathbf{r}}$ with respect to the injected spin-polarized electron state
superposed by the two eigenstates of the 2DEG in the presence of\ both Rashba
and Dresselhaus SO coupling. The RSP behaves like a windshield wiper resulting
from the conservation of $\left\langle \mathbf{S}\right\rangle _{\mathbf{r}%
,\perp}^{R}$ and the oscillation of $\left\langle \mathbf{S}\right\rangle
_{\mathbf{r},\parallel}^{R}$ as shown in Eq. (\ref{<S>R}) while the DSP owns
similar "swinging" behavior stemming from the conservation and oscillation of
either $\left\langle \mathbf{S}\right\rangle _{\mathbf{r},\parallel}^{D}$ or
$\left\langle \mathbf{S}\right\rangle _{\mathbf{r},\perp}^{D}$ only on certain
directions and thus exhibits crystallographic-direction dependence. The
general behavior of the spin precession due to Rashba and Dresselhaus terms
can therefore be envisioned as a superposition of RSP and DSP. This implies
that the RSP behavior always exists on [110] and [1\={1}0] directions, and
therefor an improvement in the quality of Datta-Das transistor may be achieved
by setting the 2DEG channel growing toward [001] with transport direction
being either [110] or [1\={1}0]. Furthermore, in both extreme cases a straight
path on which no spin precession occurs can always be found. With proper
arrangement, this zero spin precession path may be found to be the same
([1\={1}0], for example, when setting the injected spin pointing to $\pi/4$)
in both the Rashba and Dresselhaus cases. This implies an interesting
direction toward which the spin transport is precessionless even in the
presence of both Rashba and Dresselhaus SO couplings with different coupling strengths.

This work was supported by the Republic of China National Science Council
Grant No. 92-2112-M-002-050.

\end{document}